\documentclass[aps,prd,twocolumn,superscriptaddress]{revtex4}
\usepackage{natbib}
\usepackage{graphicx}
\usepackage[hypertex]{hyperref}
\usepackage{amssym,aas_macros}
\usepackage{color}

\def\beq{\begin{equation}}
\def\eeq{\end{equation}}
\def\bea{\begin{eqnarray}}
\def\eea{\end{eqnarray}}
\def\Nst{N_{\rm stream}}
\def\mst{m_{\rm stream}}

\def\neff{n_{\rm eff}}
\def\efg{{\varepsilon}}

\begin{document}

\title{Hierarchical Phase Space Structure of Dark Matter Haloes:
\\Tidal debris,
Caustics, and Dark Matter annihilation }
\author{Niayesh Afshordi}\email{nafshordi@perimeterinstitute.ca}\affiliation{Perimeter Institute
for Theoretical Physics, 31 Caroline St. N., Waterloo, ON, N2L 2Y5,Canada}
\author{Roya Mohayaee}\email{roya@iap.fr}\affiliation{Institut
d'Astrophysique de Paris, CNRS, UPMC, 98 bis boulevard Arago, France}
\author{Edmund
Bertschinger}\email{edbert@mit.edu}\affiliation{Department of
Physics and Kavli Institute for Astrophysics and Space
Research, MIT\\ Room 37-602A, 77 Massachusetts Ave., Cambridge, MA
02139, USA}

\date{\today}
\preprint{astro-ph/yymmnnn}

\begin{abstract}
Most of the mass content of dark matter haloes is expected to be in the form of {\it tidal debris}.
The density of debris is not constant, but rather can grow due to formation of {\it caustics}
at the apocenters and pericenters of the orbit, or decay as a result of phase mixing.
In the phase space, the debris assemble in a hierarchy which is truncated by the primordial temperature of dark matter.
Understanding this phase structure can be of significant importance for the interpretation of many astrophysical
observations and in particular dark matter detection experiments. With
this purpose in mind, we develop a general theoretical framework to describe the
hierarchical structure of the phase space of cold dark matter
haloes. We do not make any assumption of spherical symmetry and/or smooth and continuous accretion.
 Instead, working with correlation functions in the action-angle space, we can fully account for the
 hierarchical structure (predicting a two-point correlation
function $\propto \Delta J^{-1.6}$ in the action space), as well as the
 primordial
discreteness of the phase space. As an application, we estimate the boost to the dark matter annihilation
signal due to the structure of the phase space within virial radius: the boost due
to the hierarchical tidal debris is of order unity, whereas the primordial
discreteness of the phase structure can boost the
total annihilation signal by up to an order of magnitude. The latter is dominated by the regions
beyond 20\% of the virial radius, and is largest for the recently formed haloes with the least degree of phase mixing.
\end{abstract}

\maketitle

\section{Introduction}

Cosmological N-body simulations show that dark matter (DM) haloes
which form in a $\Lambda$CDM Universe
contain a large number of subhaloes of all sizes and masses. What remains outside the subhaloes
are ungrouped individual particles whose masses set
the resolution limit of the simulation. If the simulations where to have enough resolution
to resolve every single subhalo then it is expected that the smallest subhaloes would be
the microhaloes of about $10^{-6}\,M_\odot$ \citep{hofmann,diemand,bertschinger2}. Does all
the mass of a given halo reside inside the gravitationally bound subhaloes ?
As a subhalo falls through the gravitational field of its host halo, it becomes tidally disrupted.
A tidal stream extends along the orbit of the subhalo and can
contain a large fraction of the satellite mass.
Therefore, a significant fraction of a DM halo is expected to be in the form of {\it streams} and
{\it caustics}.
Depending on their length, the density of the streams can vary and
is relatively not very large. However, as a
stream folds back on itself, zones of higher density, {\it i.e.} caustics,
form (see {\it e.g.} \citep{sikivieipser,sikivietkachev,natarajan,duffy,2008arXiv0809.0497W}).
In principle, these are not true caustics but only smeared-out caustics due to finite
DM velocity dispersion, however, it is convenient to refer to them simply as DM caustics.
Hereafter we shall refer to unbound streams and caustics jointly as {\it tidal debris}.

Dark matter tidal debris, so far mostly unresolved in cosmological N-body simulations,
are expected to populate our own halo. Many stellar counterparts to such debris
have been detected so far ({\it e.g.} \citep{ibata,1999Natur.402...53H})
and many more are expected to be
detected with future missions like GAIA.
The hierarchical
growth of the host halo from the disruption of satellite haloes
reflects in a hierarchical structure of the phase space. The true lowest cutoff to
this hierarchy is not set by the microhaloes but by primordial dark
matter velocity dispersion.
The hierarchical phase structure indicates that after removing
all bound subhaloes from a given DM halo,
its phase space remains still unsmooth due to debris from
disrupted subhaloes. The tidal debris are never
smeared out because of conservation of phase space density and volume, although
they become less dense as they wrap around the halo.
It is this phase structure which we study here.

Secondary infall or
self-similar accretion model provides a solid theoretical
base for the study of halo formation, and models the phase structure of
DM haloes \citep{fillmoregoldreich,bertschinger}.
However, since this model assumes continuous
accretion, it cannot capture the hierarchical nature of halo
formation.
On the other hand, numerical simulations still lack enough resolution to
resolve the hierarchical phase structure, although progress is being made
in this direction \citep{Vogelsberger,Kuhlen,2008arXiv0809.0497W}.

Here, we aim at capturing the hierarchical phase structure of dark
matter haloes and its intrinsically discrete nature,
without resorting to any assumption of spherical symmetry
or smooth and self-similar accretion.
We divide the structure of a dark matter halo into three categories: (1)
the primordial and intrinsically discrete phase structure, formed prior to any merger or
accretion and entirely due to the coldness of the initial condition; (2) the
hierarchical phase structure of tidal debris from disrupted satellites, and
(3) the hierarchical phase structure of undisrupted subhaloes. We leave the study of
the undisrupted substructures to a companion paper \cite{phase_bound}
and in this work we
only study cases (1) and (2).

\begin{figure}
\includegraphics[width=\linewidth]{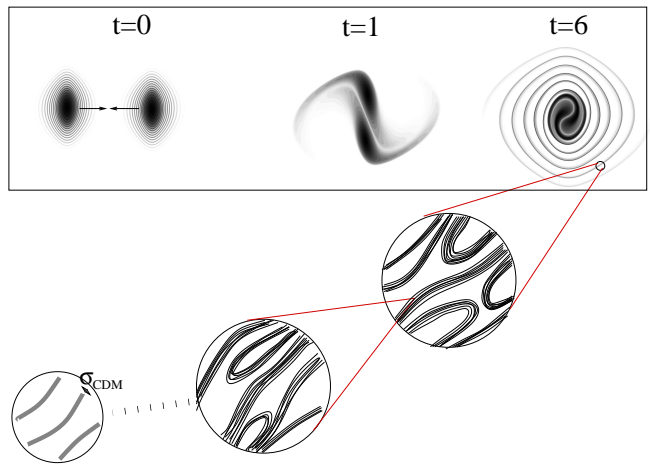}
 \caption{
The top horizontal panel shows the phase space of the merger of two dark matter haloes,
 each of which has its own hierarchy of phase structure. The times on the top panel refer to
the crossing times. The zooming shows that each hierarchy contains a
 lower level and so on. The hierarchy is cut at the scale of the smallest
 dark matter halo that has been accreted to the final halo.
However, the phase space is not smooth below this
 scale. Indeed, the phase space is intrinsically discrete due to
 the coldness of dark matter shown by the last zooming on the left.
(Top panel: courtesy of Vlasov-Poisson simulation \citep{colombi}.)
}
 \label{fig:schematic}
\end{figure}

To study phase structure induced by debris from disrupted
satellites, we assume that at a given level in the hierarchy,
all structures added earlier and which
lie at smaller scales are smooth. This sets the lowest level of
the hierarchy at the scale determined by the velocity dispersion of
earliest dark matter haloes. However, this is not entirely correct since the
earliest dark matter haloes themselves are not smooth and have a structure
which is due to the coldness of the initial condition. Thus, there is a
fundamental discreteness scale which is determined by primordial dark
matter velocity dispersion (see Fig. \ref{fig:schematic}).

This complicated process is studied here through correlation functions
in the action-angle space where Hamiltonian is only a function of the adiabatic invariants, {\it
i.e.} the action variables. Their conjugate variables, {\it the angle
variables} increase linearly in time. The action-angle variables are
extremely useful for studying tidal streams \citep{tremaine,1999MNRAS.307..495H,helmi,binney,2008MNRAS.tmp.1091F}.
However, working with the action-angle variables, we are restricted to
regions within the virial radius (with a quasi-static potential), and hence
the phase structures that might arise outside
the virial radius ({\it e.g.} between the virial and the turnaround radii)
cannot be studied in the present framework.
For direct DM detection
and cosmic-ray signal of DM annihilation, only the nearby
phase structure plays a r\^ole and our method
is valid (see {\it e.g.} \cite{mohayaeesalati}). However,
for lensing experiments and $\gamma$-ray emission from DM
annihilations, for example from other galaxies, the structures outside
the virial radius can be rather important (see {\it e.g.} \cite{Gavazzi}).

\begin{figure}
\includegraphics[width=\linewidth]{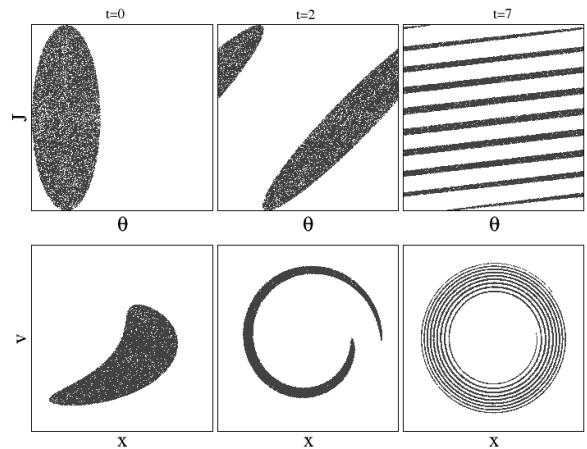}
 \caption{\textcolor{black}{A one dimensional cartoon of the evolution of tidal streams in both phase and action-angle spaces. As structure wraps around the phase space, more streams cross the same angle coordinate, which leads to a discrete lattice-like structure in the action space.}
}
 \label{fig:action-angle}
\end{figure}

We assume that the satellite orbits are integrable in the host
DM potential \textcolor{black}{(although, in \ref{sec:conclusion} we remark on chaos and non-integrable systems)}. Therefore, the phase space distribution can
be described in terms of the action-angle variables, $\{J_i,\theta_i\}$,
so that:
\bea \dot{\theta}_i = \frac{\partial {\cal H}}{\partial J_i} = \Omega_i, \\
\dot{J}_i = -\frac{\partial {\cal H}}{\partial \theta_i} = 0,
\label{eq:action-angle}
\eea
where the Hamiltonian, ${\cal H} = {\cal H}[J]$, is
only a function of action variables, ${\bf J}_i$
and $\Omega_i$s are the angular frequencies. Fig. (\ref{fig:action-angle}) shows a cartoon picture of phase mixing in the action-angle space, and its correspondence to the real space.

The hierarchical phase structure and its {\it fundamental} discreteness
set by primordial DM velocity dispersion are captured by the
correlation function of the phase density. Since, after a long time, the
distribution in the angle space is uniform, the phase density is only a
function of the action variables. This can be easily seen by writing the collisionless Boltzmann equation for the equilibrium distribution in the action-angle space:
\beq
\frac{\partial f}{\partial t} + \dot{\theta}_i \frac{\partial f}{\partial \theta_i} + \dot{J}_i \frac{\partial f}{\partial J_i} =0,
\eeq
which, combining with Eq. (\ref{eq:action-angle}), implies that the equilibrium phase space density can only be a function of action variables (and is known as the {\it strong Jeans theorem} \cite{1987gady.book.....B}).

This enables us to evaluate the density-density correlation function.
Our results are only valid statistically for typical haloes and thus may
not agree with results obtained for individual haloes in the simulations.

The nature of DM remains a mystery. Supersymmetry and
extra-dimensional extensions of the standard electroweak model provide
a natural candidate in the form of a weakly interacting and massive particle
(hereafter WIMP). These species should fill up the galactic halo.
If DM consists of WIMP's, they are expected to strongly annihilate
in the dense regions of our halo
and generate in particular gamma-rays and charged cosmic rays.
Hence, hierarchical structure of phase space can lead to the enhancement
of DM annihilation signal \citep{hogan}.
We also evaluate the boost to the annihilation signal due to
tidal debris and discreteness of the phase structure. We show that the boost
due to tidal debris is of order one, whereas
the boost from the discrete phase structure can be
up to one order of magnitude higher.

In Section \ref{sec:streams} we review a few basic relations for action-angle variables.
In Section \ref{sec:debris} and \ref{sec:discrete}, we describe the correlation functions
that would account for the phase structure
due to tidal debris and their discreteness.
In Section \ref{sec:annihilationdebris}, we evaluate the boost on
the annihilation signal due to tidal debris.
In Section \ref{sec:annihilationdiscrete} , we evaluate the
boost of the annihilation signal
from intrinsic discreteness of the phase structure, and finally
Section \ref{sec:conclusion} concludes the paper.


\section{Streams and Coherence Volume of the Phase Space}
\label{sec:streams}

\begin{figure*}
\includegraphics[width=\linewidth]{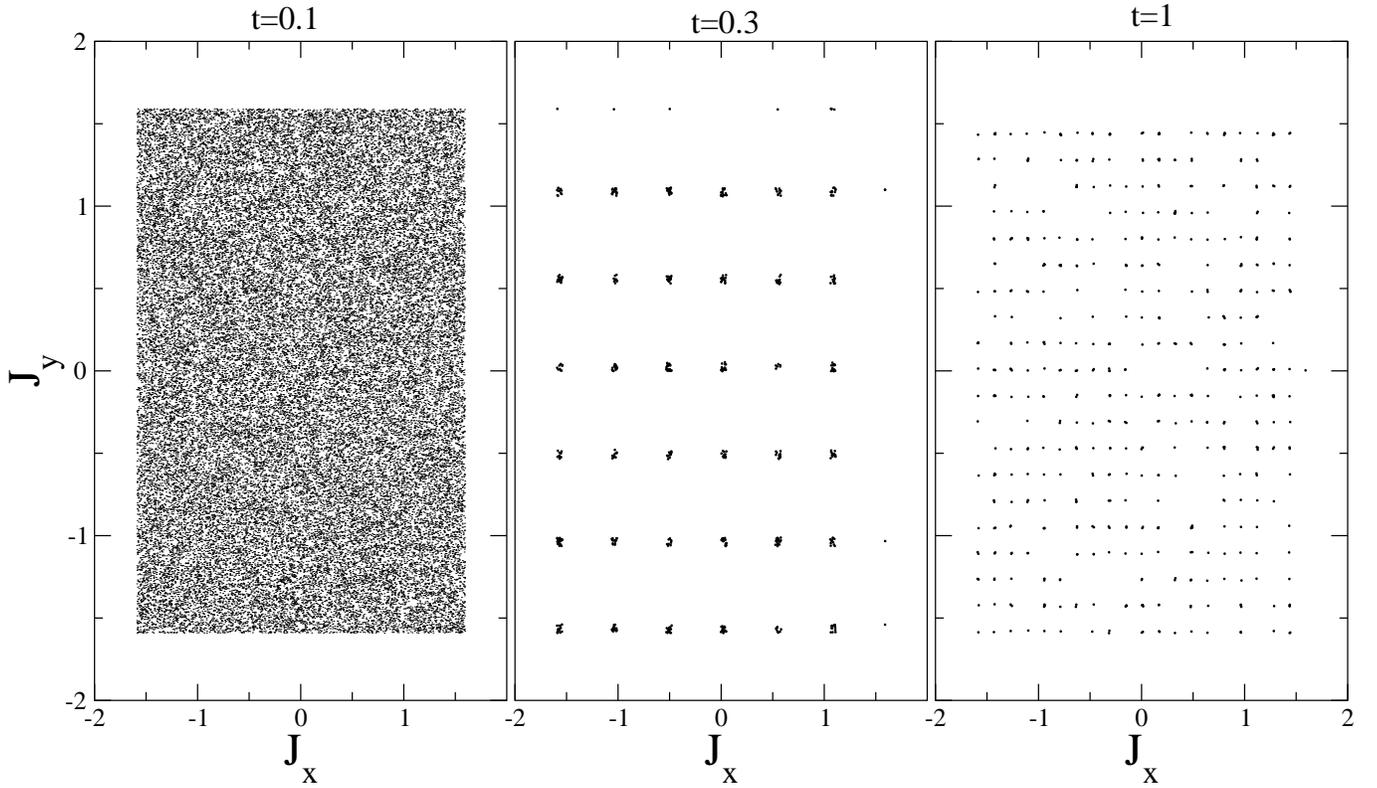}
 \caption{The action-space distribution of debris in a unit 2d torus with unit particle mass and no potential. The debris is originally within $0<x,y<0.1$, and $-10<v_x,v_y<10$. The figures show a cut through the action space with $0.09<x,y<0.1$, which is characterized by $\tilde{f}_p ({\bf J}, {\bf x}, t_{acc,p})$ (Eq. \ref{eq:tilde_f}) in our formalism.}
 \label{fig:binney}
\end{figure*}

We use the definition of action-angle variables (\ref{eq:action-angle}) and
assume that the frequencies are not degenerate, {\it i.e.}
the Hessian matrix
\beq
{\cal H}_{ij} \equiv {\partial^2 {\cal H}
\over
\partial J_i \partial J_j} = {\partial \Omega_i \over \partial J_j},
\label{hijdef}
\eeq
has non-zero eigenvalues, or equivalently, a
non-vanishing determinant:
\beq
|{\cal H}_{ij}| \neq 0,
\eeq
with the possible exception of a zero measure region of the phase space.
Note that, this implies that the halo potential cannot be assumed to
be exactly spherically symmetric, as two of the frequencies would
be equal.

If a satellite galaxy has originally a small spread in the
action variables, $\Delta J_i$, its spread in the angle variables
 increases as:
\beq
\Delta\theta_i = \left({\cal
H}_{ij} \Delta J_j \right)t_{acc,p} +\Delta\theta_0,
\label{eq:deltatheta}
\eeq
where $t_{acc,p}$ is the time since accretion of the progenitor of the debris into the halo. The last term, the initial extent of the debris, is subdominant at large times.
We set this term to zero for now, but at the end of
Section \ref{sec:annihilationdiscrete}, we discuss
when it can become important and how it could affect our results.
Therefore, the total volume swept in the angle space grows as
\beq
\Delta^3\theta = \left(|{\cal
H}_{ij}| \Delta^3J\right) t_{acc,p}^3 = (\Delta^3\Omega)~ t_{acc,p}^3\,
\label{eq:theta0}
\eeq
where we
used the definition of ${\cal H}_{ij}$ in equation (\ref{hijdef}),
and $\Delta^3\Omega$ is the volume occupied by the debris of satellite
particles in the frequency space.

As the total volume of the angle
space is $(2\pi)^3$, the number of streams passing through each
angular coordinate is
\beq
\Nst = (\Delta^3\Omega) \left(t_{acc,p}\over
2\pi\right)^3.
\label{nst}
\eeq
Thus, the total mass of each stream,
$\mst$, is the mass of the debris,
$m$, divided by $\Nst$
\beq \mst = \frac{m}{\Nst} = \left(m\over
\Delta^3\Omega\right) \left(t_{acc,p}\over 2\pi\right)^{-3}.\label{mstr}
\eeq
Put another way, the action space is divided into cells of volume:
\beq
\Delta^3J_{\rm stream} = \left(2\pi\over t_{acc,p}\right)^3
\,|{\cal H}_{ij}|^{-1}\;,
\label{d3J}
\eeq
as a result of phase space mixing ({\it e.g.} \cite{binney}) .

\textcolor{black}{
With this picture in mind, we can write the distribution in the action space as the sum of the contributions from individual progenitors:
\beq
f({\bf J},\theta)= \sum_{p} f_p({\bf J},\theta;t_{acc,p}), \label{eq:progenitor}
\eeq
where each $f_p$ has a cellular structure characterized by Eq. (\ref{d3J}), as shown in Fig. (\ref{fig:binney}), which gets finer and finer with time. Eq. (\ref{eq:progenitor}) is the phase space analog of the widely used {\it halo model} in cosmology \cite{1991ApJ...381..349S,2002PhR...372....1C}, where the density is assumed to be the sum of contributions from individual haloes with given profiles. Correspondingly, $f_p$ characterizes the profile of individual progenitors in our picture.
}

\textcolor{black}{We can now write the real space density as:
\bea
\rho({\bf x},t) = \sum_p \int d^3Jd^3\theta f_p({\bf J},\theta;t_{acc,p})\delta^3_D[{\bf x} - \tilde{\bf x}({\bf J},\theta)]\nonumber\\
=\sum_p \int d^3J \tilde{f}_p({\bf J}, {\bf x}, t_{acc,p}) \tilde{\rho}({\bf x}; {\bf J}),
\eea
where $\tilde{\rho}({\bf
x};{\bf J})$ is the density of a
distribution of unit mass, with a fixed action variable ${\bf J}$,
and uniform angle distribution:
\beq
\tilde{\rho}({\bf
x};{\bf J}) \equiv \int \frac{d^3\theta}{(2\pi)^3} \delta^3_D\left[{\bf x}-\tilde{\bf x}({\bf \theta}, {\bf J}) \right],\label{eq:density_t}
\eeq
while
\beq
\tilde{f}_p ({\bf J}, {\bf x}, t_{acc,p}) \equiv \tilde{\rho}({\bf
x};{\bf J})^{-1} \int \frac{d^3\theta}{(2\pi)^3} f_p({\bf J},{\bf \theta},t_{acc,p}) \delta^3_D\left[{\bf x}-\tilde{\bf x}({\bf \theta}, {\bf J}) \right]. \label{eq:tilde_f}
\eeq}
An example of $\tilde{f}_p$ is shown in Fig. (\ref{fig:binney}) for debris in a toy model of a unit torus. \textcolor{black}{As we will explicitly show in \ref{sec:annihilationdebris}, projecting this discrete structure in the action space of the debris into the real space leads to discrete, (nearly) singular, caustics that are only smoothed by the original velocity dispersion of the progenitor}.
\section{Clustering in the Phase Space}

\textcolor{black}{Averaging over different possible realizations of the debris within a halo, the mean phase space density can be written as an integral:
\beq
\langle \sum_p \tilde{f}_p({\bf J},{\bf x},t_{acc,p}) \rangle = \int dN_p g^{(1)}({\bf J} - {\bf J}_p,{\bf x}, t_{acc,p}),
\eeq
where $g^{(1)}$ and ${\bf J}_p$ are the profile and mean action of individual progenitors, while
\beq
dN_p \equiv dm_p d^3J_pdt_{acc,p}\frac{dn}{dm_p d^3J_pdt_{acc,p}}
\eeq
is the differential progenitor number density per units of progenitor mass, $m_p$, it action space volume $d^3J_p$, and its accretion time $t_{acc,p}$. We now follow an analogy with the cosmological halo model\cite{1991ApJ...381..349S,2002PhR...372....1C} to write the clustering in the action space as a superposition of one and two-progenitor terms:
\bea
\left\langle\sum_{p_1,p_2} \tilde{f}_{p_1}({\bf J}_1,{\bf x},t_{acc,p_1})\tilde{f}_{p_2}({\bf J}_2,{\bf x},t_{acc,p_2}) \right\rangle\nonumber\\ =
\int dN_p {\rm (1-prog.)} + \int dN_{p_1} \int dN_{p_2} {\rm (2-prog.)}.
\eea
The one-progenitor term characterizes the self-clustering of individual progenitor action-space profiles.
\beq
{\rm (1-prog.)} =  g^{(1)}({\bf J}_1 - {\bf J}_p,{\bf x}, t_{acc,p})g^{(1)}({\bf J}_2 - {\bf J}_p,{\bf x}, t_{acc,p}),
\eeq
while the two-progenitor terms characterize the correlation between phase space density at different action variables, within different progenitors:
\bea
 {\rm (2-prog.)}=  g^{(2)}({\bf J}_1-{\bf J}_{p_1}, {\bf J}_2-{\bf J}_{p_2}, {\bf x},t_{acc,p_1},t_{acc,p_2}) \nonumber\\= g^{(2)}_{\rm con.}({\bf J}_1-{\bf J}_{p_1}, {\bf J}_2-{\bf J}_{p_2}, {\bf x},t_{acc,p_1},t_{acc,p_2})\nonumber\\+g^{(1)}({\bf J}_1- {\bf J}_{p_1},{\bf x}, t_{acc,p_1})g^{(1)}({\bf J}_2 - {\bf J}_{p_2},{\bf x}, t_{acc,p_2}).\nonumber\\
\eea
In the limit that the mean actions of different progenitors are not correlated, the connected part of the (2-prog.) term goes to zero: $g^{(2)}_{\rm con.} \rightarrow 0$, and thus the two-progenitor term reduces to the correlation within the smooth halo. Note that this limit cannot be strictly realized, as due to phase space conservation, phase streams tend to avoid each other, leading to $g^{(2)}<0$ at small separations
${\bf J}_{p_1}-{\bf J}_{p_2}$. However, Liouville's theorem is not valid for coarse-grained phase space density, and thus coarse-grained progenitors can overlap in the action space.
}

\textcolor{black}{The connected part of the two-progenitor term originates from  the clustering of the initial conditions of the progenitors of the host halo, which is generally expected from the clustering of cosmological haloes. However, the structure of the one-progenitor term is more subtle: In addition to the cellular structure described in the previous section (Fig. \ref{fig:binney}), the internal structure of each progenitor prior to its accretion onto the host halo would introduce a hierarchy within each cell. In fact, in a hierarchical picture of structure formation, one expects the sub-cellular structure of the one-progenitor term to be inherited from the two-progenitor terms within progenitors prior to their accretion onto the main halo (see Fig. \ref{fig:schematic}). \textcolor{black}{The key difference between the two hierarchies, however, is that phase mixing only continues in the action space of the main halo, and (following the tidal disruption) has stopped in the action spaces of the progenitors.}
}

\textcolor{black}{For statistically self-similar initial conditions, we expect the sub-cellular and two-progenitor terms to blend into one roughly self-similar structure in the action space, although individual realizations have periodic structures with the characteristic volume given in Eq. (\ref{d3J}). We provide a scaling ansatz for this structure in \ref{sec:debris}. However, the self-similarity is cut-off by the free streaming of dark matter particles on small separations, due to their finite intrinsic velocity dispersion. This is responsible for the fundamental discreteness of the phase space distribution (see Fig. \ref{fig:schematic}), which we model in \ref{sec:discrete}.}

\subsection{hierarchical phase structure from tidal debris }
\label{sec:debris}

We first consider the phase structures due to tidal debris. Once again, we
emphasize that these are the the tidal streams that have fallen into
the gravitational field of the host halo and are no longer bound to
the original satellite.

The first level of approximation that we will use to study phase
space clustering of cold dark matter (CDM) is to assume a (statistically) hierarchical
formation history, where any trace of the cold initial conditions
has been wiped out through phase mixing. Furthermore, we ignore the
possibility of gravitationally bound structures in
this paper (see the companion paper \cite{phase_bound} on this subject).
The impact of cold initial conditions will be
addressed in subsequent sections.

Assuming {\it uniform distribution in angles} (or complete phase
mixing), the density at each point in the halo is given by:
\beq
\rho({\bf x}) = (2\pi)^3\int d^3J ~f({\bf J}) \tilde{\rho}({\bf
x};{\bf J}),
\label{eq:density}
\eeq
where $f({\bf J})$ is the phase space density,
while $\tilde{\rho}({\bf x};{\bf J})$ was defined in Eq. (\ref{eq:density_t}).
Note that in (\ref{eq:density}), the function
$\tilde{\rho}$ has the dimension of inverse volume, $1/V({\bf J})$.

We shall assume adiabatic invariance; the action remains constant as new
structures are added on larger scales. Hence,
the distribution function in the action space, $f({\bf J})$, does not change with
time, except when new structures are added due to satellites that are newly
accreted inside the virial radius.

The assumption of uniformity in the angle
space allows us to separate the effect of phase mixing from that of
hierarchical structure formation. While the former is the cause of
original caustic formation, too much phase mixing (within a fixed
potential) will eventually smooth out the real space density
distribution. The assumption of a smooth $f({\bf J})$ distribution,
implies that phase mixing is complete.

On the other hand, the effect of hierarchical structure formation is
captured in $f({\bf J})$, through the fact that structure in $f({\bf
J})$ is added on different scales, at different times in the history
of the halo. A statistical measure of this history is the two point
correlation function of the action space density. We thus
hypothesize that the correlation function:
\beq
\xi_f({\bf
J}_1,{\bf J}_2) \equiv \langle f({\bf J}_1) f({\bf J}_2) \rangle
\eeq
should be a power law for statistically self-similar initial conditions:
\beq
\langle f({\bf J}_1) f({\bf J}_2) \rangle_{\rm debris}
\simeq A |\bar{\bf J}|^{-\alpha} |{\bf J}_1-{\bf J}_2|^{-\alpha},
\eeq
as long as $|{\bf J}_1-{\bf J}_2| \ll |\bar{\bf J}|$ with
$\bar{\bf J}=({\bf J}_1+{\bf J}_2)/2$.
The form of the correlation
function uses the symmetry between ${\bf J}_1$ and ${\bf J}_2$.
\textcolor{black}{It also guarantees
that small scale structures are captured, as structures are added on different scales at different times (see Figs.~\ref{fig:schematic} for demonstration). Moreover, since actions remain constant in the adiabatic invariance approximation $dA/dt \simeq 0$ on small scales}. In other words, the correlation function $\xi_f({\bf
J}_1,{\bf J}_2)$ grows inside-out in the ${\bf J}_1,{\bf J}_2$ space.

The mean phase space density of a virialized halo, assuming a virial
overdensity of $\sim 200$, is given by:
\bea
(2\pi)^3 \Delta^3 J
\sim r^3_{vir} \sigma^3_{vir} \sim \frac{(GM)^2}{10 H} \\
\Rightarrow \langle f({\bf J}) \rangle \sim \frac{10H}{G^2
M},\label{fsc}
\eea
where $M$ is the halo virial mass. The virial
action variable is also roughly:
\beq
J_{vir} \sim r \sigma_{vir}
\sim \frac{(GM)^{2/3}}{(10 H)^{1/3}}.\label{volume}
\eeq
Given that
$M \propto a^{6/(\neff+3)}$ and $H\propto a^{-3/2}$ when perturbations grow
during the matter-dominated era, with $a$ being the cosmological scale factor, and $\neff$ the slope
of the linear power spectrum, we conclude: \beq \langle f({\bf J})
\rangle \propto J^{-{3(\neff+7)\over \neff+11}} \Rightarrow \alpha =
{3(\neff+7)\over \neff+11} \simeq 1.6 \pm 0.1, \label{alpha}\eeq
where we have assumed $\neff \simeq -2.5 \pm 0.5$ for cosmological
haloes.

\textcolor{black}{
An alternative way to derive  (\ref{alpha}) is to consider the self-similar collapse models of Fillmore \& Goldreich \cite{fillmoregoldreich}, where they calculate the actions and use adiabatic invariance to find the outcome of spherical cold secondary infall in an Einstein-de Sitter universe. For the spherical self-similar linear initial condition of
\beq
\frac{\delta M}{M}|_{\rm init.} \propto  M^{-\efg}, \label{dm_m}
\eeq
they find the action at the turn-around radius scales as
\beq
J_{ta} \propto M_{ta}^{\efg+1/3} t^{\frac{2}{9\efg}-\frac{1}{3}}, M_{ta} \propto t^{2\over 3\efg}\label{action}
\eeq
for $\efg < 2/3$, where $M_{ta}$ is the mass within the turn-around radius. Although (\ref{action}) is only for the radial action, and the two other action variables vanish due to spherical symmetry, one may imagine that for triaxial CDM haloes, the three actions would become comparable: $J_{\phi} \sim J_{\theta} \sim \epsilon J_r \sim \epsilon J_{ta}$, where $\epsilon$ characterizes the triaxiality of the halo (not to be confused with the self-similar profile index $\efg$ in (\ref{dm_m})). Therefore, eliminating time from the two equations in (\ref{action}), the phase space density, $f({\bf J})$ scales as:
\beq
f({\bf J}) \sim \frac{M_{ta}}{J_{ta}^3} \sim J^{6\over 3\efg +4} \label{fj2}.
\eeq
Assuming that the self-similar linear density profile has the same radial/mass scaling as the variance of the cosmological density fluctuations, $\sigma(M) \propto M^{-(\neff+3)/ 6}$ yields $\efg \simeq (\neff +3)/6$  ($< 2/3$ for CDM Harrison-Zel'dovich primordial power spectrum), which, plugging into (\ref{fj2}), reproduces (\ref{alpha}).
 }
\subsection{fundamental discreteness of the phase space structure}
\label{sec:discrete}

In the previous section, we considered the hierarchical addition of
tidal debris to a DM halo. However, the
hierarchy has a lower cut-off set by the velocity dispersion of the
smallest accreted satellite. Micro haloes of $\sim 10^{-6}$ solar mass could
indeed determine such a cut-off \citep{hofmann,diemand,bertschinger2}. However, this
cutoff is far above the primordial velocity
dispersion of DM itself. Therefore, the primordial velocity dispersion
introduces a {\it fundamental} discreteness in the phase structure.
\textcolor{black}{In other words, the smooth phase space distribution of the last section
ignores the discrete nature of multiple streams in the phase space due to the presence of a cut-off in the CDM hierarchy.
This discreteness shows up as a cellular or lattice structure in the action space, with a characteristic cell volume given in (\ref{d3J}) (see Figs. \ref{fig:action-angle} and \ref{fig:binney} for 1d and 2d cartoons; More realistic simulated examples are discussed in \cite{binney}). After averaging over different spacings, expected for different accretion times of different debris, the discreteness would only show up as the zero-lag of the action space correlation
function:}
\bea
\langle f({\bf J}_1) f({\bf J}_2) \rangle_{\rm dis}\, \simeq \,
\frac{\mst^2}{\Delta^3 J_{\rm stream}}\delta^3_D({\bf J}_1-{\bf
J}_2)
\label{eq:cf-discrete}
\eea
where $\mst$ and $\Delta^3J_{\rm stream}$ were defined
in equations (\ref{mstr}-\ref{d3J}), and here we have assumed a zero initial
temperature for CDM particles. A finite CDM temperature will smoothen
the delta function, as the phase space density cannot exceed its
primordial value (see Fig.~\ref{fig:schematic} for a cartoon).

\section{Example: Dark Matter Annihilation Measure}
\subsection{DM annihilation in tidal debris}
\label{sec:annihilationdebris}

In this subsection, we consider the enhancement in the expectation
value of the annihilation measure due to hierarchical structures built
in the phase space from tidal debris.

For a uniform distribution in the angle space, the expectation value
of the annihilation measure is given by:
\bea
\Phi &=& \int d^3x~ \langle\rho({\bf x})^2\rangle
\nonumber\\
&=& (2\pi)^6\int d^3J_1 d^3J_2 \xi_f({\bf J}_1,{\bf J}_2)\int d^3x
\tilde{\rho}({\bf x};{\bf J}_1) \tilde{\rho}({\bf x};{\bf J}_2).
\nonumber\\
\label{eq:emissionmeasure}
\eea
We remark that the above integral can also be relevant for the
direct detection of DM, as it quantifies the variance of the density field.

In order to investigate the impact of caustics, near the apocenters and pericenters
of the orbits, \textcolor{black}{we make a simple analogy with a one-dimensional harmonic oscillator:
\beq
H = \frac{1}{2}(p^2_x+p^2_y+p^2_z)+\frac{1}{2}\omega^2x^2,\label{eq:toy}
\eeq
 For concreteness, we also assume the other two dimensions are compact with the length $L_y$ and $L_z$, although the Hamiltonian has no explicit dependence on $y$ and $z$ coordinates. As the evolution in the three spatial directions decouple, we can simply read off three action variables from the areas of phase diagrams for each direction:
 \bea
 J_x = \frac{p^2_x}{2\omega} + \frac{1}{2}\omega x^2,\\
 J_y = \frac{L_y p_y}{2\pi}, J_z = \frac{L_z p_z}{2\pi}.
\eea
From these relations, we can find $\tilde{\rho}({\bf x};{\bf J})$ using its definition in Eq. (\ref{eq:density_t}):
\beq
\tilde{\rho}({\bf x};{\bf J}) = \frac{(x_{\rm max} - x_{\rm min})}{\pi V \sqrt{\left(x-x_{\rm min}\right) \left(x_{\rm max}-x\right)}},
\eeq
where
\bea
x_{\rm max} = -x_{\rm min} = \sqrt{2J_x/\omega},\\
V = L_y L_z (x_{\rm max} - x_{\rm min}).
\eea
}
\textcolor{black}{Notice that the square root singularity in the projection kernel $\tilde{\rho}({\bf x};{\bf J})$ is very similar to the singularity expected near CDM caustics. However, for a smooth distribution in the action-space $f({\bf J})$, the real space density $\rho({\bf x})$ is an integral over the kernel (Eq. \ref{eq:density}), which would lead to a smooth $\rho({\bf x})$. Therefore, a discrete distribution in the action space is necessary to produce caustic singularities in the real space (otherwise known as fold catastrophes or Zel'dovich pancakes). }

We then notice that the toy model of Eq. (\ref{eq:toy}) is similar to the motion in a nearly spherical potential, in the sense that the motion in one direction (x or radial) is limited by requiring constant action variables, while the two other directions (y and z, or angular directions) are compact.  Based on this analogy, we will use:
\beq
\tilde{\rho}({\bf x};{\bf J}) \sim
{[r_{\rm max}({\bf J})-r_{\rm min}({\bf J})]\over  V({\bf J}) \sqrt{[r-r_{\rm min}({\bf
J})][r_{\rm max}({\bf J})-r]}},
\label{eq:densitycaustics}
\eeq
where we have assumed an integrable nearly-spherical
potential, with small angular momentum
(and third integral), while $V({\bf J})$ is the spatial volume
occupied by the stream of action ${\bf J}$.
The radii, $r_{\rm min}$
and $r_{\rm max}$ are the minimum and maximum radii of all orbits with
the same action variable. \textcolor{black}{However, the general structure of the singularity close to boundaries does not change in other geometries.}

Using Eq. (\ref{eq:densitycaustics}), we can evaluate the $x$ integral in the emission measure (\ref{eq:emissionmeasure}). For ${\bf J}_1 \simeq {\bf J}_2$, the integral is logarithmically divergent around $r \simeq r_{\rm max}({\bf J}_1) \simeq r_{\rm max}({\bf J}_2)$ as well as $r \simeq r_{\rm min}({\bf J}_1) \simeq r_{\rm min}({\bf J}_2)$. Focussing on the outer caustic $r_{\rm max}$ we find:
\bea
\int d^3x \tilde{\rho}({\bf x};{\bf J}_1) \tilde{\rho}({\bf x};{\bf J}_2)
\simeq \frac{4\pi r^2_{\rm max} (r_{\rm max} - r_{\rm min}) }{V^2({\bf J}_1)} \nonumber\\ \times{\rm cosh}^{-1}\left|r_{\rm max}({\bf J}_1) + r_{\rm max}({\bf J}_2) \over r_{\rm max}({\bf J}_1) - r_{\rm max}({\bf J}_2) \right|
\eea

This yields:
\beq
\int
d^3x \tilde{\rho}({\bf x};{\bf J}_1) \tilde{\rho}({\bf x};{\bf J}_2)
\sim V({\bf J})^{-1} \left|\ln\left( {\rm F}^i \Delta J_i \over |\bar{{\bf J}}|\right)\right|\;,\label{eq:lnF}
\eeq
where ${\rm F}^i\sim \frac{1}{4}|\bar{{\bf J}}|\partial {\rm ln} r_{\rm max}/\partial J_i$
and can be calculated, given the gravitational potential of the host halo.
Therefore, the annihilation measure takes the form:
\beq
\Phi \sim A \int d^3\bar J |\bar{\bf J}|^{-\alpha}
\int d^3\Delta{\bf J} |\Delta{\bf J}|^{-\alpha}
\left|\ln\left({\rm F}|\Delta{\bf J}|\over |\bar{\bf J}|\right)\right|.
\eeq
We see that since $3-\alpha >0$, the integral is finite and
dominated by large $\Delta J$'s. The boost to the annihilation
signal, which is introduced by small scale clustering in the action
space ({\it i.e.} that $\alpha > 0$) is thus given by:
\bea
1 & + & B_{\rm debris} \equiv  \frac{\Phi}{\Phi_{\rm smooth}}\nonumber\\
&\simeq &
{
\int d^3\bar{\bf J}\,V(\bar{\bf J})^{-1}\,\bar{\bf J}^{-\alpha}
\int d^3\Delta{\bf J}\,\Delta{\bf J}^{-\alpha} \left|{\rm ln}
\left({{\rm F}\left|\Delta{\bf J}\right|
\over \left|\bar{\bf J}\right|}\right)\right|
\over
\int d^3\bar{\bf J}\,V(\bar{\bf J})^{-1}\,\bar{\bf J}^{-2\alpha}
\int d^3\Delta{\bf J}\, \left|{\rm ln}
\left({{\rm F}\left|\Delta{\bf J}\right|\over \left|\bar{\bf J}\right|}\right)\right|
}
\nonumber\\
& \simeq & {9\over (\alpha-3)^2}\,F^\alpha = {\cal O}(1),
\label{eq:boost-debris}
\eea
given that $F \sim \partial \ln r_{\rm max}/\partial \ln J$ is a dimensionless number of order unity.
%
%
%

Therefore, we see that the boost factor
obtained here for the tidal debris is dominated by larger
separations, as seen from expression (\ref{eq:boost-debris}) whereas our
approximation (\ref{eq:densitycaustics}) is valid for small separations.
To emphasize, (\ref{eq:densitycaustics}) is valid in the vicinity of caustics whereas the integral
(\ref{eq:boost-debris}) demonstrates that most contributions come from
large separations in the action space.

To summarize, while we predict an ${\cal O} (1)$ boost in annihilation signal due to (finite separation) clustering in action space, the main effect comes from large structures (and not caustics) which are not accurately captured in our framework. In the next section, we will address the impact of the discreteness of the phase space of CDM haloes.




\subsection{DM annihilation boost due to discreteness of
phase structures and catastrophes}
\label{sec:annihilationdiscrete}

The primordial velocity dispersion of DM induces a fundamental
discreteness in the hierarchical phase structure and can enhance the
annihilation signal. The emission measure due to this discreteness is
calculated by inserting (\ref{eq:cf-discrete}) in expression
(\ref{eq:emissionmeasure}). Summing over all streams and
subtracting the smooth part which is obtained by smoothing over the
fundamental streams, we obtain
\bea
\delta\Phi_{\rm dis} &=& \sum_{\rm stream}
m^2_{\rm stream} \nonumber\\
& \times &
\int d^3x \left[\tilde{\rho}({\bf x};{\bf J_{\rm
stream}})^2-\tilde{\rho}({\bf x};{\bf J_{\rm stream}})_{\rm
smooth}^2\right]
\nonumber\\
& \simeq &
\,K^2\, \left(t\over 2\pi\right)^{-3}
\int d^3\Omega \left(dM_{\rm halo}\over d^3\Omega\right)^2
\nonumber\\
& \times &
\int d^3x \left[\tilde{\rho}({\bf
x};{\bf J})^2-\tilde{\rho}({\bf x};{\bf J})_{\rm
smooth}^2\right],
\eea
for the enhancement of the emission measure due to discreteness
of the phase structure, where $\tilde{\rho}({\bf x};{\bf J})_{\rm
smooth}$ is the stream density, smoothed to the level that different
streams overlap, and
for the stream mass we assume:
\beq
\mst \sim K\,\frac{dM_{\rm halo}}{d^3\Omega}
\left(t\over 2\pi\right)^{-3}, \label{mstr_ap}
\eeq
where $d^3\Omega$ is the volume element in the
frequency space and factor $K$
is the ratio of the density of debris to that of the host halo: $\rho_{\rm debris}=m/v \sim  K\,\rho_{\rm halo}$.

The above expression is to be compared to the
annihilation measure for the smooth halo:
\bea
\Phi_{\rm smooth}  & = &
\int d^3\Omega \left(dM_{\rm halo}\over
d^3\Omega\right)
\nonumber
\\
& \times&
\int
d^3\Omega' \left(dM_{\rm
halo}\over d^3\Omega'\right) \int d^3x
\tilde{\rho}({\bf x};{\bf
J})\tilde{\rho}({\bf x};{\bf J'}),\nonumber\\
\eea
where ${\bf \Omega}$ and
${\bf \Omega'}$ are functions of ${\bf J}$ and ${\bf J}'$
respectively. Therefore, the boost associated with a given point in
the action space is given by:
\bea
& & B_{\rm dis}[{\bf J}]  \equiv
\frac{\delta\Phi_{\rm dis}}{\Phi_{\rm smooth}} = \nonumber\\
&& K^2\, \left({t\over 2\pi}\right)^{-3}
\frac{
\left(dM_{\rm halo}\over
d^3\Omega\right)\int d^3x \left[\tilde{\rho}({\bf x};{\bf
J})^2-\tilde{\rho}({\bf x};{\bf J})^2_{\rm smooth}\right]}{\int
d^3\Omega' \left(dM_{\rm halo}\over d^3\Omega'\right) \int d^3x
\tilde{\rho}({\bf x};{\bf J})\tilde{\rho}({\bf x};{\bf J'})}.
\nonumber\\
\label{B_dis1}
\eea

In order to estimate the boost factor, we should first approximate
the density integral $\int d^3x \tilde{\rho}({\bf x};{\bf
J})\tilde{\rho}({\bf x};{\bf J'})$. As we discussed in the previous
section for $|{\bf J}-{\bf J'}| \ll |{\bf J}|$ the integral is
dominated by the regions around the turn-around radii (or caustics)
and thus grows as
\beq
\int d^3x \tilde{\rho}({\bf x};{\bf
J})\tilde{\rho}({\bf x};{\bf J'}) \sim \frac{|\ln\left(|{\bf J}-{\bf
J'}|/ |{\bf J}|\right)|}{V({\bf J})},\label{rho_int_1}
\eeq
as seen in Eq. (\ref{eq:lnF}).
In the opposite limit, $|{\bf J}-{\bf J'}| \sim |{\bf J}|$, assuming that
the two density kernels overlap, $\tilde{\rho}({\bf x},{\bf J})$ can
be approximated as roughly constant, which yields:
\beq
\int d^3x
\tilde{\rho}({\bf x};{\bf J})\tilde{\rho}({\bf x};{\bf J'}) \sim
V^{-1}({\bf J}).
\label{rho_int_2}
\eeq
Now we note that in
(\ref{B_dis1}), the integral over $\Omega'$ (or equivalently $J'$)
in the denominator is dominated by the large values of $|{\bf
J-J'}|$, while the numerator is in the small $|{\bf J-J'}|$ regime.
Therefore, substituting from (\ref{rho_int_1}-\ref{rho_int_2}),
we find:
\bea
B_{\rm dis}[{\bf J}]\!\!\! & \simeq & \!\!\!
K^2 \left(t\over
2\pi\right)^{-3}
\left|d\ln M_{\rm halo}\over d^3\Omega\right|
\left|\ln\left({|\Delta{\bf J}_{\rm CDM}|\over |{\Delta\bf J}_{\rm
int-stream}|}\right)\right|\nonumber\\
&=&\!\!\!
K^2\frac{1}{3}\left(t\over 2\pi\right)^{-3}
\left|d\ln M_{\rm halo}\over d^3\Omega\right|
\ln\left(f_{\rm CDM}\over \langle f_{\rm halo}\rangle\right),
\nonumber\\
\label{B_dis2}
\eea
where $|\Delta{\bf J}_{\rm CDM}|$ and
$|{\Delta\bf J}_{\rm int-stream}|$ characterise the fundamental CDM
stream width and the inter-stream spacing, respectively.
The stream thickness and spacing in real space
have been calculated within the framework of
selfsimilar model \citep{shandarin}. However to obtain the ratio we use a far simpler approximation.
To get the last line of expression (\ref{B_dis2}), we have used the fact that
the volume in phase space occupied by the fundamental streams
is $(2\pi)^3\Delta^3{\bf J}_{\rm CDM}=M_{\rm halo}/f_{\rm CDM}$, while the total
volume in phase space of the entire halo is:
 $(2\pi)^3\Delta^3{\bf J}_{\rm int-stream}=M_{\rm halo}/\langle f_{\rm halo}\rangle$.

The logarithmic enhancement factor is roughly:
\beq
\frac{1}{3}\ln\left(f_{\rm CDM}\over \langle f_{\rm
halo}\rangle\right) \simeq 20 + \ln\left(m_{\chi}\over 100 ~{\rm
GeV}\right),
\label{fmchi}
\eeq
for CDM particles of mass $m_{\chi}$.
In arriving at (\ref{fmchi}) we
have used that $f_{\rm CDM}\sim \Omega_m\rho_{\rm crit}
\,(T_\chi/T_{\rm CMB})^3/\sigma_{\rm CDM}^3$ and
$\langle f_{\rm halo}\rangle \sim 10^3 \Omega_m\rho_{\rm crit}/\sigma_{\rm
vir}^3$, with $T_{\rm CMB}\sim 10^{-4}$ eV, $T_{\chi} = \frac{1}{3} m_{\chi}\sigma_{\rm CDM}^2 \sim
m_{\chi}/40$ is the CDM kinetic decoupling temperature.

In order to estimate the boost factor in (\ref{B_dis2}), we also need
to find the volume element occupied in the frequency
space, $d^3\Omega$, by a given mass element. First we should notice
that for a Keplerian potential $\varphi(r)\propto -r^{-1}$, this volume
element vanishes as three frequencies are equal, {\it i.e.}
$\Omega_r=\Omega_{\phi}=\Omega_3$, where $\Omega_3$ is the frequency
for the third integral.

For a general potential, $\varphi(r)$, we have
\beq
\frac{\Omega_r}{\Omega_\phi}-1 = \sqrt{3+{d\ln\varphi^\prime\over d\ln r}}-1,
\label{eq:freqr}
\eeq
for a nearly circular orbit and where $^\prime$ indicates first
derivative w.r.t. $r$, {\it i.e.} $\varphi^\prime=d\varphi/dr$. Moreover, the
extent in $\Omega_3$ depends on the triaxiality of the halo:
\beq
d\Omega_3 \sim \epsilon
\Omega_\phi,
\label{eq:freq3}
\eeq
where $\epsilon \sim 10\%$ characterizes the typical
triaxiality of CDM haloes. Therefore, we will approximate the volume element as
\beq
d^3\Omega \sim 4\pi\,\epsilon\Omega^2_\phi\left(\sqrt{3+{d\ln\varphi^\prime\over d\ln r}}-1\right)
d\Omega_\phi,
\label{d3Omega}
\eeq
where
\beq
\Omega^2_{\phi} \sim
\frac{G M_{\rm halo}}{r^3} = {\varphi^\prime\over r}.
\label{Omegaphi}
\eeq
Hence, for a general potential we obtain
\bea
{d\ln M_{\rm halo}\over d^3\Omega}\!\!\!\!&=&\!\!\!\!
{1\over 2\pi\epsilon}
\left(2+d\ln\varphi^\prime/d\ln r\right)\left(\varphi^\prime/r\right)^{-1/2}\nonumber\\
\!\!\!\!\! & \times &\!\!\!\!\!
\left[\left(\sqrt{3+d\ln\varphi^\prime/d\ln r}-1\right)
\left(\varphi^\prime/r-\varphi^{\prime\prime}\right)\right]^{-1}\nonumber\;,\\
\label{eq:MOmega}
\eea
To evaluate (\ref{B_dis2}) we also need to know the multiplication factor
$K$, which is the ratio of the density of
disrupted satellite to that of the host halo. This factor can be
evaluated for a general spherical potential, assuming that the
debris start with the maximum density that allows them to be tidally unbound:
\beq
K\simeq 3\left(\varphi^\prime-r\varphi^{\prime\prime}\right)
\left(2\varphi^\prime+r\varphi^{\prime\prime}\right)^{-1}
\label{eq:K}
\eeq
Putting (\ref{eq:MOmega}) and (\ref{eq:K}) in (\ref{B_dis2}) we obtain
\bea
B_{\rm dis}
&=&
36\pi^2
(\varphi^\prime-r\varphi^{\prime\prime})(2\varphi^\prime+r\varphi^{\prime\prime})^{-1}
\left(\sqrt{3+\tilde\varphi}-1\right)^{-1}
\nonumber\\
&\times &(\varphi^\prime/r)^{-3/2}\left({3\over 8\pi G \rho_{\rm crit}}\right)^{-3/2}
{1\over 3\epsilon}
\ln\left({f_{\rm CDM}\over \langle f_{\rm halo}\rangle}\right)
\nonumber\\
\label{eq:Bdiscrete}
\eea
%
where $\tilde\varphi=d(\ln\varphi^{\prime})/d({\ln r})$ and $\rho_{\rm crit}=
\frac{3H_0^2}{8\pi G}$ is the critical density of the Universe.
Moreover, we have used the fact that the product of the present-day
Hubble constant and the age of the Universe is unity ($H_0 t =
1.03\pm 0.04$), for the current concordance cosmology.

\begin{figure*}
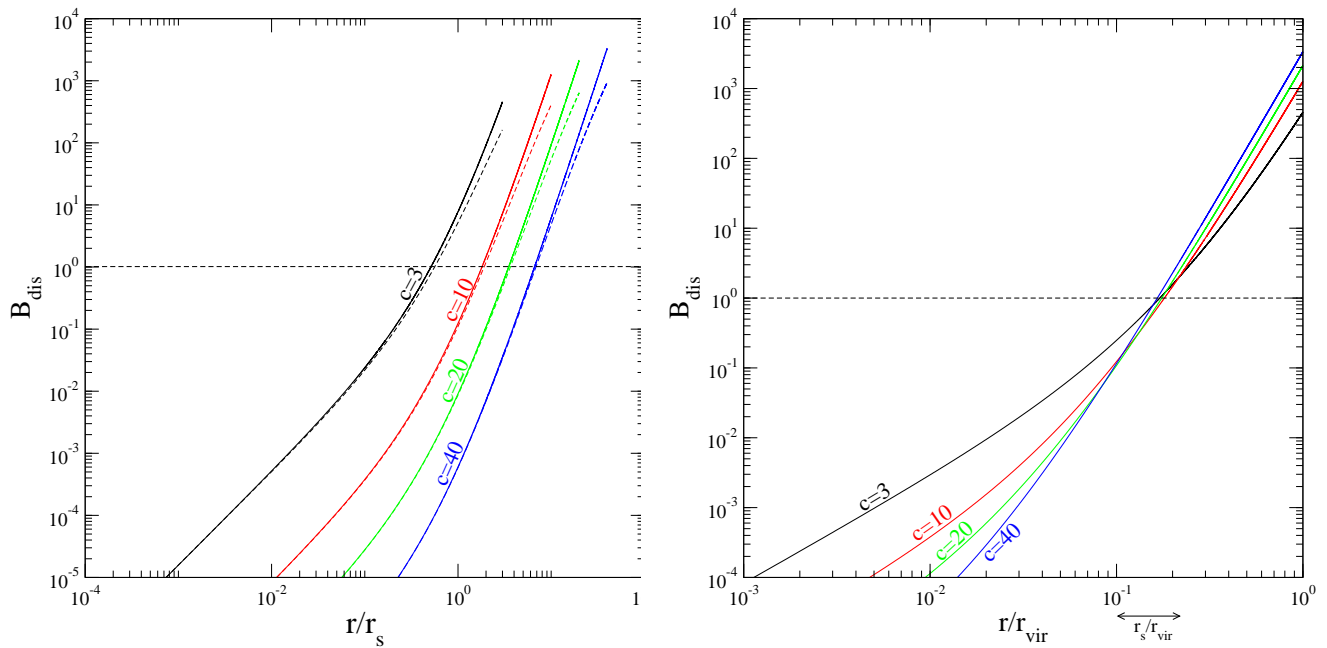

\includegraphics[width=\columnwidth]{Differential-BOOST.eps}
\includegraphics[width=\columnwidth]{Boost-NFW-y-discrete.eps}
\caption{{\it The local boost factor due to primordial discreteness of
the phase structure, for an NFW potential}:
The plots show how the boost in the annihilation measure of a
DM halo changes as we go from the inner part of halo to outer parts,
due to the discrete phase space structure of CDM.
The local boost increases as we go towards the outskirts
 of the halo and also as we decrease the concentration.
The dashed curves on the left panel show the boost if we include
 corrections
due to a finite initial phase for the debris (\ref{eq:det}), which
 become
important for nearly degenerate frequencies. The right panel shows that most of the
boost comes from regions beyond 20\% of the virial radius.
}
\label{fig:boost-NFW}
\end{figure*}


For a power-law potential:
\beq
\varphi(r) = -
\varphi_0 r^{-\beta} \qquad \beta\not=1,
\label{eq:potential}
\eeq
one can see that (\ref{eq:Bdiscrete}) yields
\beq
B_{\rm dis} =
{72\pi^2\over 3\epsilon}{(\beta+2)\sqrt{2(1-\beta)}\over (\sqrt{2-\beta}-1)}
\ln\!\!\left({f_{\rm CDM}\over \langle f_{\rm halo}\rangle}\right)
\!\!\!\left({3\rho_{\rm halo}\over \rho_{\rm crit}}\right)^{-3/2},\!\!\!\!\label{b_beta}
\eeq
where $\rho_{\rm halo}$, the local halo density for a general potential
is
\beq
\rho(r)={1\over 4\pi G}\left(\varphi^{\prime\prime}+2\varphi^\prime/r\right),
\label{eq:bdis}
\eeq
from Poisson equation.

For the power-law potential (\ref{eq:potential}), we can find the local
density in terms of the critical density and the radius and consequently obtain the boost
factor (\ref{eq:Bdiscrete}) as a function of $r/r_{\rm vir}$ where
$r_{\rm vir}$ is the virial radius of the halo, defined as the radius within which the mean density is $200 \rho_{\rm crit}$. The local boost factor for a
power-law potential, and our nominal values of $\epsilon$ and $f_{\rm CDM}/\langle f_{\rm halo}\rangle$, is
\beq
B_{\rm dis}={36\over 5}\pi^2{\beta+2\over (1-\beta)\sqrt{2-\beta}-1}
\left({r\over r_{\rm vir}}\right)^{3(\beta+2)/2}.
\eeq
The above expression shows that the boost is most significant in the outskirts of the
halo, and for large values of $\beta$. Indeed, the local boost
diverges as $\beta\rightarrow 1$. We note that, in the context of an
NFW potential\cite{nfw}:
\beq
\varphi_{\rm NFW}=-\varphi_0 {\ln(1+x)\over x},
\label{eq:nfw}
\eeq
i.e. $\beta \sim 1$ corresponds to the outskirts of the halo and
$\beta \sim -1$ to the central part.

Similarly to the power-law potential, we can obtain the local
boost factor (\ref{eq:Bdiscrete}) for an NFW potential (\ref{eq:nfw}).
The boost is shown in Fig.~\ref{fig:boost-NFW} as a function of
$r/r_s$ and for different values of the concentration parameter $c \equiv r_{\rm vir}/r_s$.
The boost increases as we go towards the outskirts of the halo
as the number of streams decreases, hence increasing the density of individual caustics. As we increase the concentration, the
central density of the halo becomes large, which in turn decreases the local boost factor.

\begin{figure}
\includegraphics[width=\linewidth]{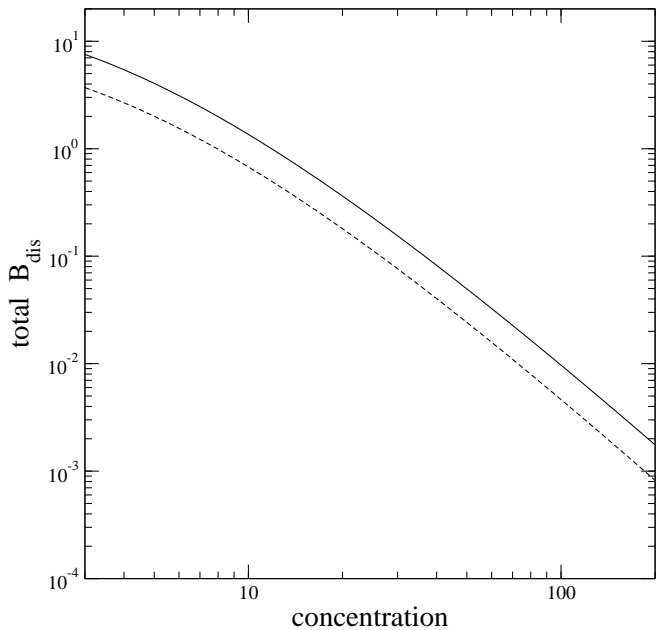}
 \caption{The estimated total boost in the annihilation measure of a
DM halo, due to the discrete distribution in the CDM phase space is shown for an NFW halo.
The lower dashed curve shows the total boost when we include corrections
 due to a finite initial phase
for the debris (\ref{eq:det}), which become important
for nearly degenerate frequencies.
 }
 \label{fig:totalboost}
\end{figure}


Having evaluated the local boost, we can evaluate the total boost from the halo
which we defined as
\beq
B_{\rm total}\equiv{\Phi_{\rm total}\over \Phi_{\rm smooth}}-1,
\label{eq:totalboost-discrete}
\eeq
where
\beq
\Phi_{\rm smooth}=r_s^3\,\rho_s^2\,\int_0^c 4\pi x^2 \rho(x)^2 dx,
\eeq
and
\beq
\Phi_{\rm total}=r_s^3\,\rho_s^2\,\int_0^c \left[1+B_{\rm dis}(x)\right]\,\rho(x)^2 4\pi x^2 dx,
\eeq
where $B_{\rm dis}$ is given by (\ref{eq:Bdiscrete}) and the density profile of the smooth halo is
given by $\rho(x)$. For NFW density profile $\rho=\rho_s\,/[x(1+x)^2]$ where $\rho_s$ is
the scale density, we have
plotted the variation of the total boost with the concentration
parameter, $c$, in Fig.~(\ref{fig:totalboost}). Again, the boost decreases
as we increase the concentration, since the flux becomes dominated by
the central part of the halo.


We saw that the boost is mostly in the outskirts of the haloes, namely beyond the $20\%$ of the
virial radius. However, one has to be cautious, since
in the outskirts of the halo the gravitational field of the halo approaches a
Keplerian potential, causing the frequencies to become
degenerate and the term $\Delta\theta_0$ in (\ref{eq:deltatheta}), which we have so far ignored,
can become important.
Thus, we need to study the importance of this term for our analysis and
 the boost.
Expression (\ref{eq:deltatheta}) now becomes
\beq
\Delta\theta=t\,\Delta\Omega\,\left(1+\frac{1}{t}{\Delta\theta_0\over\Delta\Omega}\right)\;.
\eeq
Hence the volume element in the angle space (\ref{eq:theta0}) should be replaced by
\bea
\Delta^3\theta  = (\Delta^3\Omega)~ t^3\,{\rm det}\left(
\delta_{ij}+{1\over t}
{\partial\theta_{0i}\over \partial \Omega_j}\right)\;.
\label{eq:theta02}
\eea
where det stands for the determinant.
For circular orbit approximation we have
$\Omega_\phi=\sqrt{\varphi^\prime/r}$ and we use
expressions (\ref{eq:freqr}) and (\ref{eq:freq3}) for volume element in the frequency space.
We thus find that the boost (\ref{eq:Bdiscrete}) has to be multiplied by
the inverse of the following determinant:
\bea
&{\rm det}\left(
\delta_{ij}+{1\over t}
{\partial\theta_{0i}\over \partial \Omega_j}\right)
\simeq  \left(1+{1\over t\sqrt{\varphi^\prime/r}}\right)
\left(1+{1\over \epsilon t\sqrt{\varphi^\prime/r}}\right) &
\nonumber\\
&\times \left(1+{1\over t\sqrt{\varphi^\prime/r}(\sqrt{3+\tilde\varphi}-1)}\right)&
\label{eq:det}
\eea
if the frequencies were to become near degenerate. In obtaining (\ref{eq:det}) we
have used $\Delta\theta_0/\Delta\Omega\sim 1/\Omega$ .
The effect of this factor in reducing the boost is shown for an
NFW potential (\ref{eq:nfw}) by the dashed lines
in Figs.~\ref{fig:boost-NFW} and \ref{fig:totalboost}. The local boost is
reduced slightly in the outskirts as expected and the total boost is
reduced by a factor of about 2.

\section{Conclusions}
\label{sec:conclusion}

Working in phase space and with action-angle variables, we have shown that
the density-density correlation function can capture the hierarchical phase
structure of tidal debris and also the fundamental discreteness of the phase
structure due to the coldness of the CDM initial conditions.
The study presented here assumes no spherical symmetry, no continuous or
smooth accretion, and no self-similar infall for the formation of dark
matter haloes. It is thus a general scheme
for quantifying the statistical properties of the phase structure of the virialized region of cosmological haloes.

As an application, we have obtained the significance for dark matter annihilation signal due to the
hierarchical phase structure of tidal debris and have shown that this structure
boosts the annihilation flux by order unity. On the other hand, the total boost due to
the primordial discreteness of the phase space can be one order of
magnitude higher
for low-concentration (or recently formed) dark matter haloes.

\begin{figure}
\includegraphics[width=\linewidth]{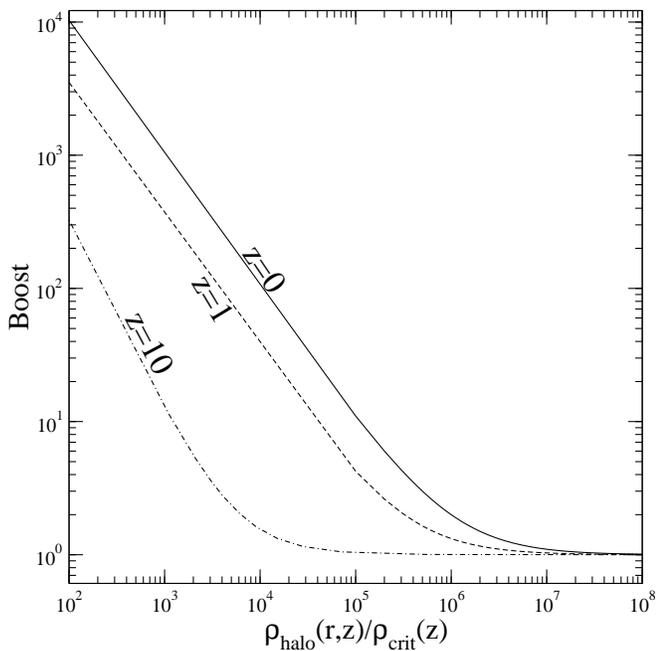}
 \caption
{The estimated local total boost including contributions from
 the debris, discreteness and the subhaloes, given
by (\ref{eq:boost-z}) with the first term set at its lowest value
 of unity, is
 shown for different redshifts. At high redshifts, the primordial caustics dominate
 over all other effects. However, at low redshifts the discreteness
 effect due to caustics is only important in the outskirts of the haloes.}
 \label{fig:boost-z}
\end{figure}

While this paper dealt with unbound debris and caustics in dark matter
haloes, in a companion paper \cite{phase_bound}, we calculate the boost
to the annihilation signal due to the {\it gravitationally bound}
substructure or sub-haloes. Combining the results of both papers, we can
write down a concise and approximate formula for
the local boost due to {\it all} substructures:
\bea
&& {\rm Boost} \equiv \frac{\langle \rho^2({\bf x}) \rangle}{\langle
\rho({\bf x}) \rangle^2}-1 = B_{\rm debris}+ B_{\rm dis} + B_{\rm sub}
\sim \nonumber\\ && {\cal O}(1) + 3\! \times \! 10^5\!
\left({\rho_{\rm crit}\over \rho_{\rm halo}({\bf
x})}\right)^{3/2}\!\!\! + 10^6 \left({\rho_{\rm crit}
\over \rho_{\rm halo}({\bf x})}\right)\!\!\left({H_0\over H}\right)^2,\nonumber\\
\label{eq:boost-z}
\eea
which should be valid within a factor of $3$ in the virialized region of
the haloes. $\rho_{\rm halo}$ is the local coarse-grained density of the
halo at redshift $z$, while $\rho_{\rm crit}$ is the critical density of the Universe at
redshift $z$. The first term in (\ref{eq:boost-z}) is due to the
hierarchical structure of CDM debris, that we estimated in Sec.
\ref{sec:annihilationdebris}.  The second
term is due to the discrete nature of the CDM phase space, where we used (\ref{b_beta})
with $\beta \sim 0.5$ and our nominal values for other parameters.
Lastly, the third term is
the contribution due to gravitationally bound sub-haloes \cite{phase_bound}.
We take $H^2/H_0^2=\Omega_m(1+z)^3+\Omega_\lambda$ and plot
(\ref{eq:boost-z}) in Fig.~\ref{fig:boost-z}.

One may wonder whether the discreteness of the phase space of sub-halos
could lead to an additional boost in the annihilation signal. In other
words, should we add $B_{\rm sub}$  and $B_{\rm dis}$ to get the total
boost, or rather should they be multiplied? To answer this question, we
notice that the main contribution to $B_{\rm sub}$ is due to the
smallest sub$^{\rm n}$-haloes (or micro-haloes) which have the highest
densities \cite{phase_bound}, while the $B_{\rm dis}$ is mainly due to
the lowest density regions of the haloes, which have the lowest degree
of phase mixing. Therefore, we expect the two terms  $B_{\rm sub}$  and
$B_{\rm dis}$ to simply add incoherently, as the cross-correlation
between the two sources of sub-structure should be small.

Finally, we should point out that the results here apply to phase
structure within virial radius, and those
outside the virial radius which we have not studied here, might yield a bigger
boost factor. It is reasonable to study the streams and caustics
that lie between the virial and the turnaround radii by using
the secondary infall model \cite{fillmoregoldreich}, as radial approximation can be reasonably applied
to regions beyond the virial radius \textcolor{black}{(see Fig.\ref{fig:boost-NFW})}.

Finally, we remark on the most instrumental assumption in our framework, which was the integrability of orbits in the CDM potential, since one cannot define action-angle variables in a non-integrable system. This is characterized by the appearance of chaotic orbits in parts of the phase space. First, we should point out that as CDM haloes have a triaxial structure, a significant fraction of halo particles cannot be on chaotic orbits. Moreover, the difference between chaotic and integrable orbits only becomes important after many orbital times, which are only possible in the inner parts of the halo. As most of the boost to the annihilation caused by the discreteness in the action space comes from the outskirts of the halo (Fig.\ref{fig:boost-NFW}), we do not expect a significant difference due to chaotic orbits. Nevertheless, the implications of chaos for the structure of the CDM phase space correlation function remains an intriguing question.

 NA is supported by Perimeter Institute for Theoretical Physics.
 Research at Perimeter Institute
is supported by the Government of Canada through Industry Canada and by
 the Province of Ontario through
the Ministry of Research \& Innovation. NA is grateful to IAP for
 hospitality. R.M. thanks French ANR
(OTARIE) for grants and Perimeter Institute for hospitality. EB acknowledges support from
NSF grant AST-0407050 and NASA grant NNG06GG99G.

\bibliographystyle{utcaps_na2}
\bibliography{phase}

\end{document}